\documentclass[twocolumn,print, pra, superscriptaddress]{revtex4}
\usepackage{amsfonts}
\usepackage{amssymb}
\usepackage{amsmath}
\usepackage{epsfig}
\usepackage{color}
\usepackage{graphics, graphicx}
\usepackage{bbold}
\usepackage{psfrag}
\usepackage{mathcomp}
\usepackage{subfigure}
\usepackage{verbatim}
\usepackage{color}
\usepackage[colorlinks,citecolor=blue]{hyperref}

{
}
\begin{document}

\title{Two-fluid theory for superfluid system with anisotropic effective masses}
\author{Yi-Cai Zhang}
\affiliation{School of Physics and Electronic Engineering, Guangzhou University,
Guangzhou 510006, China}
\author{Chao-Fei Liu}
\affiliation{School of Science, Jiangxi University of Science and Technology, Ganzhou
341000, China}
\author{Bao Xu}
\affiliation{Key Laboratory of Magnetism and Magnetic Materials at Universities of Inner
Mongolia Autonomous Region and Department of Physics Science and Technology,
Baotou Normal College, Baotou 014030, China}
\author{Gang Chen}
\email{chengang971@163.com}
\affiliation{State Key Laboratory of Quantum Optics and Quantum Optics Devices, Institute
of Laser spectroscopy, Shanxi University, Taiyuan 030006, China}
\affiliation{Collaborative Innovation Center of Extreme Optics, Shanxi University,
Taiyuan, Shanxi 030006, China}
\author{W. M. Liu}
\affiliation{Beijing National Laboratory for Condensed Matter Physics, Institute of
Physics, Chinese Academy of Sciences, Beijing 100190, China}

\begin{abstract}
In this work, we generalize the two-fluid theory to a superfluid system with
anisotropic effective masses along different principal axis directions. As a
specific example, such a theory can be applied to
spin-orbit coupled Bose-Einstein condensate (BEC) at low temperature. The normal density from phonon excitations and the second sound
velocity are obtained analytically. Near the phase transition from the plane
wave to zero-momentum phases, due to the effective mass divergence, the normal density from phonon excitation increases greatly,
while the second sound velocity is suppressed significantly. With quantum
hydrodynamic formalism, we give a unified derivation for suppressed
superfluid density and Josephson relation. At
last, the momentum distribution function and fluctuation of phase for the long
wave length are also discussed.
\end{abstract}

\maketitle

\section{Introduction}

At low temperature, Bose-Einstein condensation and superfluidity would occur
in bosonic system. Tissa \cite{Tisza} and Landau \cite{Landau}
propose two-fluid theory to explain the superfluid phenomena in Helium-4.
Comparing with usual classical fluid, due to an extra degree of freedom
(existence of condensate), the existence of second sound is an
important characteristic of superfluidity. With realizations of Bose-Einstein
condensate (BEC)  and fermion superfluidity in dilute atomic gas, the second sound and other related
superfluid phenomena in atomic gas have attracted great interests \cite{Taylor2005,He2007,Taylor2009,Hu2014,Hou2013,Meppelink2009,Tey2013}.
For
example, sound velocities at zero temperature as a function of density in
cold atoms \cite{Andrews1997,Joseph2007} have been
measured experimentally.
The application of two-fluid theory for sound
propagations in cold atomic gas has been proposed \cite%
{Zaremba1998,Shenoy1998}.
The predictions on the second sound \cite%
{Heiselberg2006,Arahata2009} and the quenched moment of
inertia \cite{Baym2013} resulting from superfluidity in cold atoms has been observed
experimentally \cite{Riedl2011, Sidorenkov2013}.
According to the two-fluid theory, the whole fluid can be viewed as a mixture of two component fluids,
namely, the normal part and superfluid part. The motions of normal part result in viscosity, while the motions of superfluid one are dissipationless.
As temperature grows from absolute zero to superfluid transition point, the superfluid density decreases from total density to zero.
Specially, the normal density at usual superfluid system (Helium-4 fluid or cold atoms) is vanishing at zero temperature. Consequently, the moment of inertia is also vanishing in usual isotropic superfluid system at zero temperature.




Recently spin-orbit coupled BEC has been realized experimentally \cite%
{Lin1,Wang2,Cheuk,zhangjinyi,Olson,Khamehchi}. There exist a phase transition between the plane wave phase and the zero momentum phase in the spin-orbit coupled BEC \cite{Lin1,zhangshizhong}.
It is shown that, even at zero temperature,
there exists finite
normal density, and even all the total density becomes normal at the phase transition point although
the condensate fraction is finite \cite{normaldensity}. At zero
temperature, due to finite normal density, there is finite momentum of
inertia in the spin-orbit coupled BEC \cite{stringari2016}. It is shown
that the suppressing of superfluid density is closely related to
enhancements of effective masses near the ground state. Because the effective masses enhance anisotropically, the expansion behaviors of spin-orbit
coupled gas also shows anisotropy \cite{Martone2012,Qu2017,zhangyongping}.

It is expected that due to enhancements of effective masses in spin-orbit
coupled BEC, the corresponding two-fluid theory at finite temperature
also need to be revised greatly. In this work, we generalize the two-fluid theory to a superfluid system with
anisotropic effective masses along different principal axis directions. As an immediate application, we find that a lot of superfluid
properties of spin-orbit coupled BEC, e.g., the decreasing of superfluid
density, the suppressed anisotropic sound velocities, etc., can be described
by an anisotropic two-fluid theory.
Near the phase transition from the plane wave to zero-momentum phases, the normal density from phonon excitation
increases greatly, while the second sound velocity is suppressed
significantly.


 The paper is organized as follows.
In Sec.~II, we review the thermodynamic relations for superfluid system. In
Sec.~III, based on the entropy equation, we give a derivation for
dissipationless two-fluid equations. In Sec.~IV, as an
application of the anisotropic two-fluid theory, we give a
specific example, namely, spin-orbit coupled BEC, to illustrate the above
results. A summary is given in Sec.~V.

\section{Thermodynamic relations for superfluid system}

First of all, we consider an original system $K_{0}$ with the particle mass $m$,
in which the many-particle Hamiltonian
\begin{equation}
H_{0}=\sum_{i}\frac{\mathbf{p}_{0i}^{2}}{2m}+\frac{1}{2}\sum_{i\neq j}V(%
\mathbf{r}_{i}-\mathbf{r}_{j}),
\end{equation}%
where $\mathbf{p}_{0i}$ is the  particle momentum for $K_{0}$ and $V(\mathbf{r}_{i}-%
\mathbf{r}_{j})$ is the interaction potential between particles $i$ and $j$. In the
following, we mainly investigate the effects arising from enhancements of
the effective masses, i.e., $m\rightarrow zm$ with $z>1$. For this purpose, we
consider another system $K$ with the effective mass $m^{\prime }=zm$. The
corresponding Hamiltonian and Lagrangian are written as
\begin{align*}
&H=\sum_{i}\frac{\mathbf{p}_{i}^{2}}{2zm}+\frac{1}{2}\sum_{i\neq j}V(\mathbf{r%
}_{i}-\mathbf{r}_{j}),\\ 
&L=\sum_{i}\frac{zm\mathbf{\tilde{v}}_{i}^{2}}{2}-\frac{1}{2}\sum_{i\neq j}V(\mathbf{r%
}_{i}-\mathbf{r}_{j}),  
\end{align*}%
where $\mathbf{p}_{i}$ and $\mathbf{\tilde{v}}_{i}$ are the particle momentum and velocity
for $K$, respectively.
From the Hamilton's canonical equations (or the Newton's second law), i.e., $d\mathbf{p}_{0i}/d t=-\partial V/\partial \mathbf{r}_{i}$ and $d\mathbf{p}_{i}/d t=-\partial V/\partial \mathbf{r}_{i}$,
 and the relations $\mathbf{p}_{0i}=m\mathbf{v}_{0i}$, $%
\mathbf{p}_{i}=zm\mathbf{\tilde{v}}_{i}$, we get the velocity for $K$ in terms of that of $%
K_{0}$, i.e.,
\begin{equation}
\mathbf{\tilde{v}_{i}}=\mathbf{v}_{0i}/z,  \label{velocityrelation}
\end{equation}%
where $\mathbf{v}_{0i}$ is the particle velocity for $K_{0}$ with the mass $m$.
Equation (\ref{velocityrelation}) shows that the enhancements of masses would
result in the decreasing of velocity. In the following, the velocity
appearing in expressions is always referred to that of the original
system $K_{0}$, which has the mass $m$, rather $zm$. The Lagrangian for $K$ can
also be expressed in terms of $\mathbf{v}_{0i}$, i.e.,
\begin{equation*}
L=\sum_{i}\frac{m\mathbf{v}_{0i}^{2}}{2z}-\frac{1}{2}\sum_{i\neq j}V(\mathbf{%
r}_{i}-\mathbf{r}_{j}).
\end{equation*}

In order to get the thermodynamic relations, now we consider a moving reference frame with the velocity $\mathbf{%
u}$ with respect to the laboratory reference frame. The particle velocity in the moving frame is
\begin{equation}
\mathbf{v}^{\prime}_{i}=\mathbf{v}_{0i}-\mathbf{u}.
\end{equation}%
 The Lagrangian $L$ is rewritten as
\begin{equation*}
L=\sum_{i}\frac{m(\mathbf{v}{^{\prime }}_{i}+\mathbf{u})^{2}}{2z}-\frac{1}{2}%
\sum_{i\neq j}V(\mathbf{r}_{i}-\mathbf{r}_{j}).
\end{equation*}%
The canonical momentum and Hamiltonian in the moving frame are thus given respectively by
\begin{equation*}
\mathbf{p}^{\prime }_{i}\equiv \frac{\partial L}{\partial \mathbf{v}^{\prime }}_{i}%
=m(\mathbf{v}^{\prime }_{i}+\mathbf{u})/z=\mathbf{p}_{0i}/z=\mathbf{p}_{i}/z.
\end{equation*}
\begin{equation*}
H^{\prime }\equiv \sum_{i}\mathbf{p}_{i}^{\prime }\cdot \mathbf{v}%
_{i}^{\prime }-L=H-\frac{\mathbf{u}}{z}\cdot \mathbf{P},
\end{equation*}%
where the total momentum $\mathbf{P}=\sum_{i}\mathbf{p}_{i}$.

In terms of the Hamiltonian $H'$, the partition function
\begin{equation}
Z\equiv tre^{-\beta H^{\prime }}=e^{-\beta F}=e^{-\beta \lbrack E-TS-%
\mathbf{u}\cdot \mathbf{P}/z]},  \label{trace}
\end{equation}%
where $E$ is the energy in the laboratory frame, $S$ is the entropy, $%
\beta =1/T$ is the inverse temperature, and the free energy
\begin{equation}
F=E-TS-\mathbf{u}\cdot \mathbf{P}/z.  \label{thermo}
\end{equation}%
The grand potential
\begin{equation*}
\Omega \equiv -pV=F-\mu N=E-TS-\mathbf{u}\cdot \mathbf{P}/z-\mu N,
\end{equation*}%
where $p$ is the pressure, $V$ is the system volume, $\mu $ is the chemical
potential, and $N$ is the total particle number. Further introducing the energy
density $\epsilon =E/V$, the entropy density $s=S/V$, the momentum density $%
\mathbf{g}=\mathbf{P}/V$, and the particle number density $n=N/V$, the
pressure is given by
\begin{equation}
p=-\epsilon +Ts+\mathbf{u}\cdot \mathbf{g}/z+\mu n.  \label{pressure}
\end{equation}

Since the free energy is a function of $\left\{ T,V,\mathbf{u},N\right\} $,
e.g., $F=F(T,V,\mathbf{u},N)$, using Eq.~(\ref{thermo}), we obtain
\begin{align}
& dF=-SdT-pdV+\mu dN-\mathbf{P}\cdot d\mathbf{u}/z  \notag
\label{freeenergyrelation} \\
& =dE-TdS-SdT-\mathbf{u}\cdot d\mathbf{P}/z-\mathbf{P}\cdot d\mathbf{u}/z,
\end{align}%
which leads to the fundamental thermodynamic relation
\begin{equation}
TdS=dE+pdV-\mu dN-\mathbf{u}\cdot d\mathbf{P}/z.  \label{TF}
\end{equation}%
For a fixed unit volume ($dV\equiv 0$), Eq.~(\ref{TF}) turns
into
\begin{equation}
Tds=d\epsilon -\mu dn-m\mathbf{u}\cdot d\mathbf{j},
\end{equation}%
where $\mathbf{j}\equiv \mathbf{g}/(zm)$ is the particle current density.


On the other hand, using $\mathbf{p}_{i}^{2}/(2zm)-\mathbf{u}\cdot
\mathbf{p}_{i}/z=(\mathbf{p}_{i}-m\mathbf{u})^{2}/(2zm)-m\mathbf{u}^{2}/(2z)$%
, Eq.~(\ref{trace}) becomes
\begin{equation*}
Z=e^{\beta Nm\mathbf{u}^{2}/(2z)}Z_{0}\equiv e^{\beta Nm\mathbf{u}%
^{2}/(2z)}tre^{-\beta H},
\end{equation*}%
where $Z_{0}=tre^{-\beta H}\equiv e^{-\beta F_{0}}$ and
 $F_{0}$ is the free energy when the fluid is at rest. So the free energy
\begin{equation} \label{equation1}
F=F_{0}-Nm\mathbf{u}^{2}/(2z).
\end{equation}
For superfluid system, Eq.~(\ref{equation1}) can be extended to a case in which the superfluid and normal parts move
with the velocities $\mathbf{v}_{s}=\hbar \mathbf{\nabla}\theta /m$
and $\mathbf{v}_{n}=\mathbf{u}$, respectively \cite{Chaikin}, where $\theta $ is the phase of the condensate order parameter. In this case, the free
energy density, $f=F/V$, is given by
\begin{equation}\label{freeenergy}
f=f_{0}-nm\mathbf{v}_{n}^{2}/(2z)+n_{s}m(\mathbf{v}_{s}-\mathbf{v}%
_{n})^{2}/(2z),
\end{equation}%
where $f_{0}$ is the free energy density when the fluid is at rest. The term $%
n_{s}m(\mathbf{v}_{s}-\mathbf{v}_{n})^{2}/(2z)$ describes an extra energy due
to the motion of the superfluid part relative to the normal part, and $n_{s}$
is the particle number density of the superfluid part. We should remind that
the velocity for $K$ is $\mathbf{\tilde{v}}_{s(n)}=\mathbf{v}_{s(n)}/z$.


The free energy density $f$ is a function of independent variables $\{T,n,\mathbf{v}_{n},\mathbf{v}_{s}\}$.
Similarly as Eq.~(\ref{freeenergyrelation}), its variation can be written as
\begin{equation}\label{free}
df=-sdT+\mu dn-m\mathbf{j}\cdot d\mathbf{v}_{n}+\mathbf{h}\cdot d\mathbf{v}%
_{s},
\end{equation}
where $\mathbf{h}\equiv\partial f/\partial \mathbf{v}_{s}$ is the thermodynamic conjugate variable of $\mathbf{{v}_{s}}$.
From Eqs.~(\ref{freeenergy}) and (\ref{free}), the particle current density and the conjugate variable of $\mathbf{v}%
_{s}$ are given respectively by
\begin{align}
& \mathbf{j}=-\frac{\partial f}{m\partial \mathbf{v}_{n}}=\frac{n_{n}\mathbf{%
v}_{n}+n_{s}\mathbf{v}_{s}}{z},  \notag  \label{h} \\
& \mathbf{h}=\frac{\partial f}{\partial \mathbf{v}_{s}}=\frac{n_{s}m(\mathbf{%
v}_{s}-\mathbf{v}_{n})}{z},
\end{align}%
where $n_{n}\equiv n-n_{s}$ is the particle number density of the normal
part. From Eqs.~(\ref{freeenergyrelation}) and (\ref{h}), the thermodynamic
relations are generalized as
\begin{align}
& p=-\epsilon +Ts+m\mathbf{v}_{n}\cdot \mathbf{j}+\mu n,  \notag
\label{relation0} \\
& Tds=d\epsilon -\mu dn-m\mathbf{v}_{n}\cdot d\mathbf{j}-\mathbf{h}\cdot d%
\mathbf{v}_{s},\notag\\
& dp= sdT +nd\mu +m\mathbf{j} \cdot d\mathbf{v}_{n} -\mathbf{h}\cdot d%
\mathbf{v}_{s}.
\end{align}%
Equation (\ref{relation0}) also holds for the anisotropic superfluid system.

\section{two-fluid equations for anisotropic effective masses}
Having obtained the fundamental thermodynamic relations in Eqs.~(\ref{h}) and (\ref{relation0}), in this section we extend them to derive the required two-fluid equations for anisotropic effective masses.
For an anisotropic system with different effective masses along three
 principal axis directions, the Hamiltonian
\begin{align}
& H=H_{0}+H_{\text{int}},  \notag \\
& H_{0}=\int d^{3}\mathbf{r}\psi ^{\dagger }\left( \frac{p_{x}^{2}}{2m_{1}}+%
\frac{p_{y}^{2}}{2m_{2}}+\frac{p_{z}^{2}}{2m_{3}}\right) \psi ,  \notag \\
& H_{\text{int}}=\frac{1}{2}\int d^{3}\mathbf{r}_{1}d^{3}\mathbf{r}_{2}\psi
^{\dagger }(\mathbf{r}_{1})\psi ^{\dagger }(\mathbf{r}_{2})V(\mathbf{r}_{1}-%
\mathbf{r}_{2})\psi (\mathbf{r}_{2})\psi (\mathbf{r}_{1}),  \label{TH}
\end{align}%
where $m_{i}$ is the effective mass along the $i$th axis and $\psi $
is the bosonic field operator. We should note that although the masses are anisotropic, the Hamiltonian (\ref{TH}) still has Galilean
transformation invariance \cite{Hou2015}, and can describe the spin-orbit
coupled BEC near the ground state realized in recent experiments \cite{Lin1}.
In specific, we write the effective mass as
\begin{equation*}
m_{i}=mz_{i},
\end{equation*}%
where $z_{i=1,2,3}\geq 1$ characterize the enhancements of masses.

\subsection{Two-fluid equations}

To obtain the two-fluid equations for the Hamiltonian (\ref{TH}), we generalize the free energy density in Eq.~(\ref{freeenergy}) as
\begin{equation}
f=f_{0}(T,n)-\sum_{i=1,2,3}\frac{nmv_{ni}^{2}}{2z_{i}}+\sum_{i=1,2,3}\frac{%
n_{s}m(v_{si}-v_{ni})^{2}}{2z_{i}}.  \label{GFE}
\end{equation}
Based on Eqs.~(\ref{h}) and (\ref{GFE}), the particle current density and the conjugate variable of the superfluid velocity  of the $i$th axis are given respectively by
\begin{align}
&j_{i}=-\frac{\partial f}{m\partial v_{ni}}=\frac{n_{n}v_{ni}+n_{s}v_{si}}{%
z_{i}}, \notag\\
&h_{i}=\frac{\partial f}{\partial v_{si}}=\frac{n_{s}m(v_{si}-v_{ni})}{z_{i}}.
\end{align}

Although there exists anisotropy, the particle number, momentum and energy
are still conserved. The corresponding continuity equations are
given respectively by
\begin{align}
& \frac{\partial n}{\partial t}+\sum_{i}\partial _{i}j_{i}=0,
\label{continuity} \\
& \frac{\partial g_{i}}{\partial t}+\sum_{j}\partial _{j}\pi _{ij}=0, \\
& \frac{\partial \epsilon }{\partial t}+\sum_{i}\partial _{i}j_{i}^{\epsilon
}=0,\label{econ}
\end{align}%
where $g_{i}=m_{i}j_{i}=z_{i}mj_{i}$, $\pi _{ij}$ is the pressure
tensor and $j_{i}^{\epsilon }$ is the
energy current density. The superfluid velocity can be written as a gradient of
condensate phase, i.e., $\mathbf{v}_{s}=\hbar \mathbf{\nabla} \theta /m$. Therefore,
the superfluid velocity $\mathbf{v}_{s}$ is irrotational and satisfies the
equation \cite{fluid}
\begin{equation}\label{vs}
\frac{m\partial v_{si}}{\partial t}+\partial _{i}(\mu +X)=0,
\end{equation}%
where $\mu $ is the chemical potential and $X$ is a scalar function which
need to be determined by an entropy equation (see the following). The
irrotationality condition is
\begin{equation}\label{irrotional}
\partial _{i}v_{sj}=\partial _{j}v_{si}.
\end{equation}%
We should note that the superfluid velocity for the anisotropic system with the
mass $z_{i}m$, i.e., $\tilde{v}_{si}=v_{si}/z_{i}$ [see Eq.~(\ref%
{velocityrelation})]  would have no irrotationality \cite%
{stringari2016} due to $z_{i}\neq z_{j}$ in general.

The entropy equation can be derived as follows. Using the thermodynamic relations in Eq.~(\ref{relation0}),
continuity equations~(\ref{continuity})-(\ref{econ}), Eqs.~(\ref{vs}) and (\ref%
{irrotional}), we get
\begin{align}
& T\left[\frac{\partial s}{\partial t}+\sum_{i}\partial _{i}\left(\frac{sv_{ni}}{z_{i}}%
+\frac{Q_{i}}{T}\right)\right]  \notag  \label{entropy} \\
& =-\sum_{i}Q_{i}\frac{\partial _{i}T}{T}  \notag \\
& -\sum_{i}\left(\frac{g_{i}}{z_{i}m}-\frac{nv_{ni}}{z_{i}}-\frac{h_{i}}{m}%
\right)\partial _{i}\mu   \notag \\
& -\sum_{ij}\left(\frac{\pi _{ji}}{z_{j}}-\frac{p}{z_{i}}\delta _{ij}-\frac{%
mj_{j}v_{ni}}{z_{i}}-\frac{v_{sj}h_{i}}{z_{j}}\right)\partial _{i}v_{nj}  \notag \\
& -\sum_{i}\left(\frac{X}{m}-\sum_{j}\frac{v_{sj}v_{nj}}{z_{j}}\right)\partial
_{i}h_{i}.
\end{align}%
In deriving  Eq.~(\ref{entropy}), we have introduced the heat current density $\mathbf{Q}$ with
\begin{align*}
& Q_{i}\equiv j_{i}^{\epsilon }-\frac{\mu (g_{i}/m-nv_{ni})}{z_{i}}-\sum_{j}\frac{%
v_{nj}\pi _{ji}}{z_{j}}-\frac{\epsilon v_{ni}}{z_{i}} \\
& +\sum_{j}\frac{mv_{ni}v_{nj}j_{j}}{z_{i}}+\left(\sum_{j}\frac{v_{nj}v_{sj}}{%
z_{j}}-\frac{X}{m}\right)h_{i},
\end{align*}%
and used the thermodynamic relation $p=-\epsilon +Ts+m\mathbf{v}_{n}\mathbf{j}%
+\mu n$. The right-hand side of Eq.~(\ref{entropy}) is a form of
\textquotedblleft currents" time \textquotedblleft forces" for entropy
production. For dissipationless process, the entropy production should be zero,
so the right-hand side should vanish, i.e.,
\begin{align}\label{constitutive}
& Q_{i}=0,  \notag \\
& \frac{g_{i}}{z_{i}m}-\frac{nv_{ni}}{z_{i}}-\frac{h_{i}}{m}=0,  \notag \\
& \frac{\pi _{ji}}{z_{j}}-\frac{p\delta _{ij}}{z_{i}}-\frac{mj_{j}v_{ni}}{%
z_{i}}-\frac{v_{sj}h_{i}}{z_{j}}=0,  \notag \\
& \frac{X}{m}-\sum_{j}\frac{v_{sj}v_{nj}}{z_{j}}=0.
\end{align}

From Eq.~(\ref{constitutive}), we get constitutive relations
\begin{align}\label{eq25}
& X=\sum_{j}\frac{mv_{sj}v_{nj}}{z_{j}},  \notag \\
& g_{i}=mnv_{ni}+z_{i}h_{i}=n_{n}mv_{ni}+n_{s}mv_{si}=z_{i}mj_{i},  \notag \\
& \pi _{ji}=p\delta _{ij}+\frac{z_{j}mj_{j}v_{ni}}{z_{i}}+v_{sj}h_{i},
\notag \\
& j_{i}^{\epsilon }=\frac{\mu (g_{i}/m-nv_{ni})}{z_{i}}+\frac{v_{nj}\pi _{ji}%
}{z_{j}}+\frac{\epsilon v_{ni}}{z_{i}}-\frac{mv_{ni}v_{nj}j_{j}}{z_{i}}.
\end{align}%

Due to Eq.~(\ref{eq25}), the entropy equation (\ref{entropy}) becomes its conservation equation
\begin{equation}
\frac{\partial s}{\partial t}+\sum_{i}\partial _{i}\left(\frac{sv_{ni}}{z_{i}}\right)=0.
\end{equation}%
The energy conservation equation can be replaced by the entropy conservation
equation. Finally, we have four complete equations for the two-fluid theory
\begin{align}
& \frac{\partial n}{\partial t}+\sum_{i}\partial _{i}j_{i}=0,  \label{twofluid1}
\\
& \frac{\partial g_{i}}{\partial t}+\sum_{j}\partial _{j}\pi _{ij}=0,\label{two3} \\
& \frac{\partial s}{\partial t}+\sum_{i}\partial _{i}\left(\frac{sv_{ni}}{z_{i}}%
\right)=0, \label{two2} \\
& \frac{m\partial v_{si}}{\partial t}+\partial _{i}\left(\mu +\sum_{j}\frac{%
mv_{sj}v_{nj}}{z_{j}}\right)=0,  \label{twofluid2}
\end{align}%
with constitutive relations
\begin{align*}
& j_{i}=\frac{n_{n}v_{ni}+n_{s}v_{si}}{z_{i}}, \\
& g_{i}=z_{i}mj_{i}=mn_{n}v_{ni}+mn_{s}v_{si}, \\
& \pi _{ji}=p\delta _{ij}+\frac{mn_{n}v_{nj}v_{ni}+mn_{s}v_{sj}v_{si}}{z_{i}}%
.
\end{align*}%
Equations~(\ref{twofluid1})-(\ref{twofluid2}) are the main results of this paper. These equations have several important properties. Firstly,  due to $z_{i}\neq z_{j}$
in general, the pressure tensor $\pi _{ij}$ would not be a symmetrical
tensor in the anisotropic case, i.e., $\pi _{ij}\neq \pi _{ji}$.

Secondly, when $z_{1}=z_{2}=z_{3}=1$, using the relation between the energy ($\epsilon $) of
the laboratory frame and that ($\epsilon _{0}$) of another
reference frame where the superfluid part is at rest \cite{fluid}, i.e.,
$\epsilon =nm\mathbf{v}_{s}^{2}/2+\mathbf{g}_{0}\cdot \mathbf{v}_{s}+\epsilon
_{0}$
with $\mathbf{g}_{0}=n_{n}m(\mathbf{v}_{n}-\mathbf{v}_{s})$, and further
comparing the thermodynamic relation in Eq.~(\ref{relation0}) with its counterpart in \cite%
{fluid}, i.e., $
d\epsilon _{0}=Tds+\mu _{0}dn+(\mathbf{v}_{n}-\mathbf{v}_{s})\cdot d\mathbf{g%
}_{0}$,
we immediately get the relation for two chemical potentials $\mu $ and $\mu
_{0}$, i.e.,
$\mu _{0}+m\mathbf{v}_{s}^{2}/2=\mu +m\mathbf{v}_{s}\cdot \mathbf{v}%
_{n}$.
Here $\mu _{0}=\partial \epsilon _{0}/\partial n$ denotes the chemical
potential for the reference frame in which the superfluid part is at
rest, while $\mu =\partial \epsilon /\partial n$ is the chemical potential
for the laboratory frame. Using replacement of $\mu +m%
\mathbf{v}_{s}\cdot \mathbf{v}_{n}\rightarrow \mu _{0}+m\mathbf{v}_{s}^{2}/2$
in Eq.~(\ref{twofluid2}), Eqs.~(\ref%
{twofluid1})-(\ref{twofluid2}) recover the famous Landau-Khalatnikov's
two-fluid equations \cite{Khalatnikov} with constitutive relations $%
j_{i}=n_{n}v_{ni}+n_{s}v_{si}$, $g_{i}=mj_{i}$, and $\pi _{ji}=p\delta
_{ij}+mn_{n}v_{nj}v_{ni}+mn_{s}v_{sj}v_{si}$. For the anisotropic case, the
relation between two chemical potentials is given by
\begin{equation}
\mu _{0}+\sum_{j}\frac{mv_{sj}^{2}}{2z_{j}}=\mu +\sum_{j}\frac{mv_{sj}v_{nj}%
}{z_{j}}.
\end{equation}%

Thirdly, at zero temperature ($n_{s}=n,n_{n}=0,s=0$, $v_{n}=0$), the entropy in Eq.~(%
\ref{two2}) can be neglected and the constitutive relations become $%
j_{i}=nv_{si}/z_{i}$, $g_{i}=nmv_{si}$, and $\pi _{ji}=p\delta
_{ij}+mnv_{sj}v_{si}/z_{i}$. Using the thermodynamic relation in Eq.~(\ref%
{relation0}) (Gibbs-Duhem relation for superfluid system at $T=0$), i.e., $%
dp=nd\mu -\mathbf{h}\cdot d\mathbf{v}_{s}$ and irrotational condition $%
\partial _{i}v_{sj}=\partial _{j}v_{si}$, one can show that  Eqs.~(\ref{two3}) and (\ref{twofluid2}) are equivalent.
Taking $\mu _{0}+\sum_{j}mv_{sj}^{2}/\left( 2z_{j}\right) =\mu +\sum_{j}%
mv_{sj}v_{nj}/z_{j}$ ($\mathbf{v}_{n}=0$) into account, the
two-fluid equations ({\ref{twofluid1})-(\ref{twofluid2}}) are reduced to
\begin{align}
& \frac{\partial n}{\partial t}+\sum_{i}\partial _{i}j_{i}=0,  \notag
\label{twofluid4} \\
& \frac{m\partial v_{si}}{\partial t}+\partial _{i}\left( \mu _{0}+\sum_{j}%
\frac{mv_{sj}^{2}}{2z_{j}}\right) =0,
\end{align}%
which are consistent with Eqs.~(8)-(10) for hydrodynamics of spin-orbit coupled
BEC in Ref.~\cite{Qu2017}, with replacements of $\mu
_{0}\rightarrow gn+V_{ext}$ and $v_{si}\rightarrow z_{i}\tilde{v}_{si}$ (replaced by
the velocities of $K$ [see Eq.~(\ref%
{velocityrelation})]). Therefore, in this sense, we can use the Hamiltonian of anisotropic effective mass [Eq.~(\ref{TH})] to describe the dynamics
of the spin-orbit coupled BEC near the ground state.


\subsection{First and second sounds}

It is known that the existence of second sound is an important character for
superfluidity. With the two-fluid equations~(\ref%
{twofluid1})-(\ref{twofluid2}), we can investigate the sound
propagations for the anisotropic system. If the amplitudes of sound
oscillations and the velocity fields $v_{s(n)}$ are small,
we can neglect the second order terms of velocities in the two-fluid
equations, i.e.,
\begin{align}
& \frac{\partial n}{\partial t}+\sum_{i}\partial _{i}j_{i}=0,  \notag
\label{linearequation} \\
& \frac{\partial g_{i}}{\partial t}+\partial _{i}p=0,  \notag \\
& \frac{\partial s}{\partial t}+\sum_{i}\partial _{i}\left( \frac{sv_{ni}}{%
z_{i}}\right) =0,  \notag \\
& m\frac{\partial v_{si}}{\partial t}+\partial _{i}\mu =0,
\end{align}%
with $g_{i}=z_{i}mj_{i}=mn_{n}v_{ni}+mn_{s}v_{si}$.

From the first two equations, we get
\begin{equation*}
\frac{\partial ^{2}n}{\partial t^{2}}=\sum_{i}\frac{\partial _{i}^{2}p}{%
z_{i}m}.
\end{equation*}%
From equation $g_{i}=mn_{n}v_{ni}+mn_{s}v_{si}$, we get $v_{ni}=\left(
g_{i}-mn_{s}v_{si}\right) /\left( mn_{n}\right) $ and
\begin{align*}
& \frac{\partial s}{\partial t}\simeq \sum_{i}\frac{-s}{z_{i}mn_{n}}\left(
\partial _{i}g_{i}-mn_{s}\partial _{i}v_{si}\right) , \\
& \frac{\partial ^{2}s}{\partial t^{2}}=\sum_{i}\frac{s}{z_{i}mn_{n}}\left(
\partial _{i}^{2}p-n_{s}\partial _{i}^{2}\mu \right) .
\end{align*}

By introducing the entropy for the unit mass, i.e., $\tilde{s}=s/(nm)$, and $ds=m%
\tilde{s}dn+nmd\tilde{s}$, we get
\begin{align*}
& m\tilde{s}\frac{\partial ^{2}n}{\partial t^{2}}+nm\frac{\partial ^{2}%
\tilde{s}}{\partial t^{2}}=\sum_{i}\frac{\tilde{s}}{z_{i}}\partial
_{i}^{2}p+nm\frac{\partial ^{2}\tilde{s}}{\partial t^{2}} \\
& =\sum_{i}\frac{\tilde{s}(n_{n}+n_{s})}{z_{i}n_{n}}\left( \partial
_{i}^{2}p-n_{s}\partial _{i}^{2}\mu \right) .
\end{align*}%
Using the thermodynamic relation (Gibbs-Duhem relation) $dp=nd\mu +sdT$ and $%
n=n_{s}+n_{n}$, we get
\begin{align*}
& nm\frac{\partial ^{2}\tilde{s}}{\partial t^{2}}=\sum_{i}\left[ \frac{n_{s}%
\tilde{s}}{z_{i}n_{n}}\left( n\partial _{i}^{2}\mu +s\partial
_{i}^{2}T\right) -\frac{nn_{s}\tilde{s}}{z_{i}n_{n}}\partial _{i}^{2}\mu %
\right]  \\
& =\sum_{i}\frac{n_{s}\tilde{s}}{z_{i}n_{n}}s\partial _{i}^{2}T=\sum_{i}%
\frac{nn_{s}m\tilde{s}^{2}}{z_{i}n_{n}}\partial _{i}^{2}T.
\end{align*}%
Therefore, we obtain
\begin{align}
& \frac{\partial ^{2}\tilde{s}}{\partial t^{2}}=\sum_{i}\frac{n_{s}\tilde{s}%
^{2}}{z_{i}n_{n}}\partial _{i}^{2}T.  \notag  \label{soundvelocity} \\
& \frac{\partial ^{2}n}{\partial t^{2}}=\sum_{i}\frac{\partial _{i}^{2}p}{%
z_{i}m}.
\end{align}
Equation~(\ref{soundvelocity}) describes the sound propagations with small amplitudes.

In order to solve Eq.~(\ref{soundvelocity}), we choose $(n,\tilde{s})$ as independent variables, e.g.,
\begin{align*}
& dp=\frac{\partial p}{\partial n}|_{\tilde{s}}dn+\frac{\partial p}{\partial
\tilde{s}}|_{n}d\tilde{s}, \\
& dT=\frac{\partial T}{\partial n}|_{\tilde{s}}dn+\frac{\partial T}{\partial
\tilde{s}}|_{n}d\tilde{s}.
\end{align*}%
If the sound oscillations have the plane wave forms, i.e.,
\begin{equation*}
\left(
\begin{array}{c}
\delta \tilde{s} \\
\delta n%
\end{array}%
\right) =\left(
\begin{array}{c}
A \\
B%
\end{array}%
\right) e^{i(\mathbf{q}\cdot \mathbf{r}-\omega t)},
\end{equation*}%
substituting  them into Eq.~(\ref{soundvelocity}), we get
\begin{equation}
\omega ^{2}\left(
\begin{array}{c}
A \\
B%
\end{array}%
\right) =\left(
\begin{array}{cc}
W(\alpha ,\phi )\left( \frac{\partial T}{\partial \tilde{s}}\right) _{n} &
W(\alpha ,\phi )\left( \frac{\partial T}{\partial n}\right) _{_{\tilde{s}}}
\\
\frac{1}{mZ(\alpha ,\phi )}\left( \frac{\partial p}{\partial \tilde{s}}%
\right) _{n} & \frac{1}{mZ(\alpha ,\phi )}\left( \frac{\partial p}{\partial n%
}\right) _{_{\tilde{s}}}%
\end{array}%
\right) \left(
\begin{array}{c}
A \\
B%
\end{array}%
\right) q^{2},  \label{sound}
\end{equation}%
where $\mathbf{q}=q\left[ \cos (\alpha ),\sin (\alpha )\cos (\phi ),\sin
(\alpha )\sin (\phi )\right] $ and
\begin{align}
& \frac{1}{Z(\alpha ,\phi )}=\frac{\cos ^{2}(\alpha )}{z_{1}}+\frac{\sin
^{2}(\alpha )\cos ^{2}(\phi )}{z_{2}}+\frac{\sin ^{2}(\alpha )\sin ^{2}(\phi
)}{z_{3}},  \notag \\
& W(\alpha ,\phi )=\frac{n_{s}\tilde{s}^{2}}{n_{n}Z(\alpha ,\phi )}.
\end{align}%

The
existence of non-trivial solutions in Eq.~(\ref{sound}) requires
\begin{equation*}
\text{Det}\left[ \left(
\begin{array}{cc}
W(\alpha ,\phi )(\frac{\partial T}{\partial \tilde{s}})_{n}-c^{2} & W(\alpha
,\phi )(\frac{\partial T}{\partial n})_{\tilde{s}} \\
\frac{1}{mZ(\alpha ,\phi )}(\frac{\partial p}{\partial \tilde{s}})_{n} &
\frac{1}{mZ(\alpha ,\phi )}(\frac{\partial p}{\partial n})_{\tilde{s}}-c^{2}%
\end{array}%
\right) \right] =0,
\end{equation*}
where $c=\sqrt{\omega ^{2}/q^{2}}$ is the sound velocity.
Further introducing the specific heat capacity at constant volume $C_{V}=T\left( \frac{\partial \tilde{s}}{%
\partial T}\right) _{V}$ and using relation $\frac{\partial }{\partial n}=-%
\frac{N}{n^{2}}\frac{\partial }{\partial V}$, we get $\left( \frac{\partial T%
}{\partial \tilde{s}}\right) _{n}\left( \frac{\partial p}{\partial n}\right)
_{\tilde{s}}-\left( \frac{\partial T}{\partial n}\right) _{\tilde{s}}\left(
\frac{\partial p}{\partial \tilde{s}}\right) _{n}=-\frac{N}{n^{2}}\frac{%
\partial (T,p)}{\partial (\tilde{s},V)}=\frac{T}{C_{V}}\left( \frac{\partial
p}{\partial n}\right) _{T}$, where $\frac{%
\partial (T,p)}{\partial (\tilde{s},V)}\equiv (\frac{\partial T}{\partial \tilde{s}})_V(\frac{\partial p}{\partial V})_{\tilde{s}}-(\frac{\partial T}{\partial V})_{\tilde{s}}(\frac{\partial p}{\partial \tilde{s}})_V$ is the Jacobian determinant.  Thence, the sound velocity
equation becomes
\begin{equation}
c^{4}-\left[ \frac{TW}{C_{V}}+\frac{1}{Z}\left( \frac{\partial p}{\partial
\rho }\right) _{_{\tilde{s}}}\right] c^{2}+\frac{TW}{C_{V}Z}\left(\frac{\partial p%
}{\partial \rho }\right)_{T}=0,  \label{soundequation}
\end{equation}%
where $\partial p/\partial \rho =\partial p/\left( m\partial
n\right) $ is the  compressibility.

 From Eq. (\ref{soundequation}), we can get the first sound
velocity $c_{1}$ and the second sound velocity $c_{2}$ \cite{Pitaevskii}.
We see that due to ¡Ì$\sqrt{1/Z(\alpha ,\phi )}\leq 1$,  the enhancements of effective masses would result in the decreasing of the sound velocities.

At zero temperature ($s=0$, $n_{n}=0$, $n_{s}=n$, $\mathbf{v}_{n}=0$),
the linear equation (\ref{linearequation}) is reduced to
\begin{align}
& \frac{\partial n}{\partial t}+\frac{n\partial _{x}v_{sx}}{z_{1}}+\frac{%
n\partial _{y}v_{sy}}{z_{2}}+\frac{n\partial _{z}v_{sz}}{z_{3}}=0,  \notag
\label{linearzero} \\
& m\frac{\partial v_{si}}{\partial t}+\partial_{i}\mu =0.
\end{align}%
The sound velocity $c(\mathbf{q})=c_0\sqrt{%
q_{x}^{2}/z_{1}+q_{y}^{2}/z_{2}+q_{z}^{2}/z_{3}}/q$ with $c_0=\sqrt{\partial p/\partial\rho}=\sqrt{n\partial\mu/(m\partial n)}$.

The first and second sounds may be probed by measuring the density response
function. In order to get it, we need to add an
external perturbation potential $\delta Ue^{i(\mathbf{q}\cdot \mathbf{r}%
-\omega t)}$ in Eq.~(\ref{soundvelocity}) of the sound
propagations, e.g.,
\begin{align}\label{withexternal}
& \frac{\partial ^{2}\tilde{s}}{\partial t^{2}}=\sum_{i}\frac{n_{s}\tilde{s}%
^{2}}{z_{i}n_{n}}\partial _{i}^{2}T,  \notag \\
& \frac{\partial ^{2}n}{\partial t^{2}}=\sum_{i}\frac{1}{z_{i}m}\partial
_{i}^{2}\left[p+n\delta Ue^{i(\mathbf{q}\cdot \mathbf{r}-\omega t)}\right].
\end{align}
The density response function is defined as
\begin{equation*}
\chi (\mathbf{q},\omega )=\frac{\delta n}{\delta Ue^{i(\mathbf{q}\cdot
\mathbf{r}-\omega t)}}.
\end{equation*}%
Similarly, if the solutions also have the plane wave forms, Eq.~(\ref{withexternal}) becomes
\begin{align}\label{eq40}
& \omega ^{2}\left(
\begin{array}{c}
\delta \tilde{s} \\
\delta n%
\end{array}%
\right) =\left(
\begin{array}{cc}
W(\alpha ,\phi )\left(\frac{\partial T}{\partial \tilde{s}}\right)_{n} & W(\alpha ,\phi )%
\left(\frac{\partial T}{\partial n}\right)_{\tilde{s}} \\
\frac{1}{mZ(\alpha ,\phi )}\left(\frac{\partial p}{\partial \tilde{s}}\right)_{n} & \frac{1}{%
mZ(\alpha ,\phi )}\left(\frac{\partial p}{\partial n}\right)_{\tilde{s}}%
\end{array}%
\right) \left(
\begin{array}{c}
\delta \tilde{s} \\
\delta n%
\end{array}%
\right) q^{2}  \notag \\
& +\left(
\begin{array}{c}
0 \\
\frac{n\delta U}{mZ(\alpha ,\phi )}%
\end{array}%
\right) q^{2}e^{i(\mathbf{q}\cdot \mathbf{r}-\omega t)}.
\end{align}%
From Eq.~(\ref{eq40}),  we get
\begin{equation*}
\delta n=\frac{n\left[ \omega ^{2}q^{2}-q^{4}W(\alpha ,\phi )\left(\frac{\partial T%
}{\partial \tilde{s}}\right)_{n}\right] }{mZ(\alpha ,\phi )[\omega ^{4}-(
c_{1}^{2}+c_{2}^{2}) \omega ^{2}q^{2}+c_{1}^{2}c_{2}^{2}q^{4}]}\delta
Ue^{i(\mathbf{q}\cdot \mathbf{r}-\omega t)}.
\end{equation*}%
So the density response function
\begin{align}\label{densityresponse}
& \chi (\mathbf{q},\omega )=\frac{n\left[ \omega ^{2}q^{2}-q^{4}W(\alpha
,\phi )\left(\frac{\partial T}{\partial \tilde{s}}\right)_{n}\right] }{mZ(\alpha ,\phi )\left[
\omega ^{4}-\left( c_{1}^{2}+c_{2}^{2}\right) \omega
^{2}q^{2}+c_{1}^{2}c_{2}^{2}q^{4}\right] }\notag \\
& =\frac{nw_{1}q}{2mc_{1}}\left( \frac{1}{\omega -c_{1}q}-\frac{1}{\omega
+c_{1}q}\right) \notag \\
& +\frac{nw_{2}q}{2mc_{2}}\left( \frac{1}{\omega -c_{2}q}-\frac{1}{%
\omega +c_{2}q}\right) .
\end{align}%
In Eq.~(\ref{densityresponse}), $w_{1(2)}$ is the weight  for the
first (second) sound in the density response function and
satisfies %
\begin{align}
& w_{1}+w_{2}=\frac{1}{Z(\alpha ,\phi )},  \notag  \label{weight} \\
& \frac{w_{1}}{c_{1}^{2}}+\frac{w_{2}}{c_{2}^{2}}=\frac{1}{(\frac{\partial p%
}{\partial \rho })_{T}}.
\end{align}%
Equation (\ref{weight}) shows that in the anisotropic superfluid system, the weights of sound
oscillations  decrease due to the enhancements of
effective masses.

The imaginary part of the density response function is
\begin{align}\label{eq43}
& \chi ^{\prime \prime }(\mathbf{q},\omega )=\text{Im}\left[\chi (q,\omega +i0)\right]
\notag \\
& =-\frac{\pi n}{2m}\left\{ \frac{w_{1}q}{c_{1}}\left[\delta (\omega
-c_{1}q)-\delta (\omega +c_{1}q)\right]\right.   \notag \\
& +\left. \frac{w_{2}q}{c_{2}}\left[\delta (\omega -c_{2}q)-\delta (\omega
+c_{2}q)\right]\right\} .
\end{align}%
The $f$-sum rule and the compressibility sum rules (for unit volume) \cite%
{Pines,Hu} are obtained by
\begin{align}\label{eq44}
& -\frac{1}{\pi }\int_{-\infty }^{\infty }d\omega \omega \chi ^{\prime
\prime }(\mathbf{q},\omega )=\frac{nq^{2}}{mZ(\alpha ,\phi )},  \notag \\
& \lim_{q\rightarrow 0}\left\{ -\frac{1}{\pi }\int_{-\infty }^{\infty
}d\omega \frac{\chi ^{\prime \prime }(\mathbf{q},\omega )}{\omega }\right\} =%
\frac{n}{\left( \frac{\partial p}{\partial \rho }\right) _{T}},
\end{align}%
or in terms of the dynamic structure factor $S(\mathbf{q},\omega )=\frac{-1}{%
\pi (1-e^{-\omega /T})}\mathtt{Im}[\chi (\mathbf{q},\omega +i0)]$,
\begin{align}\label{eq45}
& \int_{-\infty }^{\infty }d\omega \omega S(q,\omega )=\frac{nq^{2}}{%
2mZ(\alpha ,\phi )},  \notag \\
& \lim_{q\rightarrow 0}\int_{-\infty }^{\infty }d\omega \frac{S(q,\omega )}{%
\omega }=\frac{n}{2\left( \frac{\partial p}{\partial \rho }\right) _{T}}.
\end{align}%
Based on Eqs.(\ref{densityresponse})-(\ref{eq45}), the first and second sounds may be detected experimentally by measuring the density response function \cite{Arahata2009,Lingham}.

\subsection{Normal density and sound velocities}
Near zero temperature, the gapless phonon excitations would dominate the
thermodynamics.   In this case, the normal density and sound velocities can be obtained analytically.
 The normal density can be
calculated from phonon excitations by using the Landau's theory \cite%
{Atkins1959}. We assume that a thin tube filled with liquid moves with the
velocity $u$ along the $i$th axis direction. The normal part also moves due
to dragging by the tube and in equilibrium with tube wall, while the
superfluid part is at rest. The current associated normal part is given by
\begin{equation}\label{current}
j_{i}=\sum_{\mathbf{q}}q_{i}n(\mathbf{q}),
\end{equation}%
where $q_{i=x,y,z}$ is the $i$th component of vector $\mathbf{q}$, $n(%
\mathbf{q})=1/\left[{e^{\frac{\omega (\mathbf{q})-uq_{i}}{T}}-1}\right]$ is the Bose
distribution for phonon, the phonon energy $\omega (\mathbf{q})=c(\mathbf{q})q$, the
sound velocity $c(\mathbf{q})=c_{0}\sqrt{%
q_{x}^{2}/z_{1}+q_{y}^{2}/z_{2}+q_{z}^{2}/z_{3}}/q$, and $c_{0}=\sqrt{\partial
p/\partial \rho }$ is the sound velocity determined by the compressibility at
zero temperature. The average drift velocity of the phonon gas is exactly
given by
\begin{equation}\label{mean}
\bar{v}=\frac{\sum_{\mathbf{q}}v_{i}n(\mathbf{q})}{\sum_{\mathbf{q}}n(%
\mathbf{q})}=u,
\end{equation}%
with the phonon group velocity $v_{i}=\partial \omega (\mathbf{q})/\partial
q_{i}$.

On the other hand, the current from the normal part is given by $%
j_{i}=\rho _{ni}\bar{v}$ with the normal density $\rho _{ni}$. From Eqs. ~(\ref{current}) and (\ref{mean}) and taking the limit of $u\rightarrow 0$, we get
\begin{equation}
\rho _{ni}=z_{i}\sqrt{z_{1}z_{2}z_{3}}\frac{2\pi ^{2}T^{4}}{45\hbar
^{3}c_{0}^{5}}.  \label{normaldensity}
\end{equation}%
Equation~(\ref{normaldensity}) shows that the normal density satisfies the relation $\rho _{nx}:\rho _{ny}:\rho
_{nz}=z_{1}:z_{2}:z_{3}$. When $z_{i}=1$, the normal density is reduced to
the Landau's result $\rho _{n,Landau}=2\pi ^{2}T^{4}/\left( 45\hbar
^{3}c_{0}^{5}\right) $ \cite{Landau,Pitaevskii}. The correction of the
normal density relative to the usual Landau's result is given by
\begin{equation*}
\beta _{i}\equiv \rho _{ni}/\rho _{n,Landau}=z_{i}\sqrt{z_{1}z_{2}z_{3}}.
\end{equation*}%
The normal particle number density
\begin{equation}\label{eq49}
n_{n}=\rho _{ni}/(z_{i}m)=\sqrt{z_{1}z_{2}z_{3}}\frac{2\pi ^{2}T^{4}}{%
45m\hbar ^{3}c_{0}^{5}}.
\end{equation}%
Equation~(\ref{eq49}) shows that when the effective masses increase, the normal density from
phonon excitations also increases. This is because that when $z_{i}\geq1$, the
phonon excitation energy $\omega _{q}$ decreases for a fixed momentum $q$,
then the phonon number also increases for a given temperature $T$.

Near zero temperature, the free energy is given by
\begin{align*}
& F=E_{0}+F_{\text{phonon}}, \\
& F_{\text{phonon}}=Vf_{\text{phonon}}=-T\sum_{q}\text{ln}\left[\frac{1}{1-e^{-\omega (q)/T}}\right] \\
& =\frac{TV}{(2\pi \hbar )^{3}}\int d^{3}\mathbf{q}\text{ln}\left[1-e^{-\omega
(q)/T}\right]=-\sqrt{z_{1}z_{2}z_{3}}\frac{V\pi ^{2}T^{4}}{90\hbar ^{3}c_{0}^{3}},
\end{align*}
where $E_{0}$ is the ground state energy. The entropy and heat capacity are
given respectively by
\begin{align*}
& \tilde{s}=-\frac{\partial f_{\text{phonon}}}{nm\partial T}=\sqrt{z_{1}z_{2}z_{3}}\frac{%
2\pi ^{2}T^{3}}{45nm\hbar ^{3}c_{0}^{3}}, \\
& C_{V}=T\frac{\partial \tilde{s}}{\partial T}=\sqrt{z_{1}z_{2}z_{3}}\frac{%
2\pi ^{2}T^{3}}{15nm\hbar ^{3}c_{0}^{3}}.
\end{align*}
The adiabatic compressibility would equal the isothermal
compressibility, i.e., $(\frac{\partial p}{\partial \rho })_{\tilde{s}}\simeq (\frac{%
\partial p}{\partial \rho })_{T}$, so we get the first and second sound velocities from Eq.~(\ref{soundequation}) as
\begin{align}
& c_{1}=\sqrt{\frac{1}{Z(\alpha ,\phi )}\left( \frac{\partial p}{\partial
\rho }\right) }=\sqrt{\frac{1}{Z(\alpha ,\phi )}}c_{0},  \notag
\label{secondsound} \\
& c_{2}=\sqrt{\frac{TW(\alpha ,\phi )}{C_{V}}}=\frac{1}{%
(z_{1}z_{2}z_{3})^{1/4}}\frac{c_{1}}{\sqrt{3}}.
\end{align}%
For isotropic system ($z_1=z_2=z_3=1$), the above formula [Eq.~(\ref{secondsound})] for second sound recovers the famous
Landau's result, i.e., $c_{2}=c_{1}/\sqrt{3}$ \cite{Landau}. Comparing with the
usual case, the first sound velocity is suppressed by a factor $\sqrt{%
1/Z(\alpha ,\phi )}$; while the second sound velocity is suppressed by a factor $\sqrt{%
1/(Z(\alpha ,\phi )\sqrt{z_{1}z_{2}z_{3}})}$. The correction of the second sound
along the $i$th axis direction is given by
\begin{equation*}
\gamma _{i}\equiv \frac{c_{2}}{(c_{0}/\sqrt{3})}=\sqrt{1/(z_{i}\sqrt{%
z_{1}z_{2}z_{3}})}.
\end{equation*}%
As $T\rightarrow 0$, the weight of second sound in the density response functions is
proportional to difference between two compressibility, i.e., $\Delta (\frac{%
\partial p}{\partial \rho })\equiv (\frac{\partial p}{\partial \rho })_{%
\tilde{s}}-(\frac{\partial p}{\partial \rho })_{T}\propto T^{4}\rightarrow 0$%
. So, the weight of first sound $w_{1}\rightarrow 1/Z(\alpha ,\phi )$, while the
weight for second sound $w_{2}\rightarrow 0$ [see Eq.~(\ref{weight})].
We note that the normal density and sound velocities in dipolar superfluid bosons with anisotropic interactions also have been investigated \cite{Pastukhov1,Pastukhov2}.

\section{Spin-orbit coupled BEC}
In this section, we would take spin-orbit coupled BEC as example to illustrate above discussions. The corresponding Hamiltonian is given by \cite%
{Wang2,Cheuk,Olson,zhangjinyi,Khamehchi}
\begin{align}
& H=H_{0}+H_{\text{int}},  \notag  \label{eqn2} \\
& H_{0}=\int d^{3}\textbf{r}\psi ^{\dagger }\left[ \frac{(p_{x}-k_{0}\sigma
_{z})^{2}+p_{y}^{2}+p_{z}^{2}}{2m}+\frac{\Omega }{2}\sigma _{x}\right] \psi ,
\notag \\
& H_{\text{int}}=\frac{1}{2}\int d^{3}\mathbf{r}[g\psi _{1}^{\dagger }(%
\mathbf{r})\psi _{1}^{\dagger }(\mathbf{r})\psi _{1}(\mathbf{r})\psi _{1}(%
\mathbf{r})  \notag \\
& +2g^{\prime }\psi _{1}^{\dagger }(\mathbf{r})\psi _{2}^{\dagger }(\mathbf{r%
})\psi _{2}(\mathbf{r})\psi _{1}(\mathbf{r})  \notag \\
& +g\psi _{2}^{\dagger }(\mathbf{r})\psi _{2}^{\dagger }(\mathbf{r})\psi
_{2}(\mathbf{r})\psi _{2}(\mathbf{r})],
\end{align}%
where $k_{0}$ and $\Omega $ are the strengths of the spin-orbit and Raman
couplings, respectively. $\psi _{1(2)}$ is the boson field operator and $\psi ^{\dagger }=[\psi _{1}^{\dagger },\psi _{2}^{\dagger }]$ is
 the spinor form. $g=4\pi \hbar ^{2}a_{s}/m$ and $g\prime $ $=4\pi \hbar
^{2}a_{s}^{\prime }/m$ are the strengths of the intra- and inter-species
interactions with $a_{s}$ and $a_{s}^{\prime }$ being the s-wave scattering
lengths.  The above Hamiltonian breaks the Galilean transformation invariance \cite{zhuqizhong}, however we will see that the effective low energy hydrodynamics for sound oscillations restore the Galilean invariance \cite{Hou2015}. In the following, we
focus on the case of the U(2) invariant interaction, i.e., $g^{\prime }=g$, and
 set $m=1$ and $\hbar =1$ for simplicity.

 At zero temperature, the mean-field ground state wave function of the Hamiltonian~(\ref{eqn2}) is written as
\cite{zhangshizhong,zhengwei,Yip,Zhai,liyun,Martone2012}
\begin{equation*}
|0\rangle =\sqrt{n_{0}}\left(
\begin{array}{c}
\cos (\theta ) \\
-\sin (\theta )%
\end{array}%
\right) e^{ip_{0}x},
\end{equation*}%
where $n_{0}$ is the atom number density in
condensates. For weakly interacting boson gas, $n_{0}\approx n$ (the total
particle number density).
When $\Omega <2k_{0}^{2}$, $p_{0}=k_{0}\sqrt{1-\Omega ^{2}/\left(
4k_{0}^{4}\right) }$ and $\cos (2\theta )=p_{0}/k_{0}$; while for $\Omega
>2k_{0}^{2}$, $p_{0}=0$ and $\theta =\pi /4$. A quantum phase transition
occurs at $\Omega =2k_{0}^{2}$ where the sound velocity along the $x$-axis direction
becomes zero \cite{zhengwei,Ji,Khamehchi}.

\subsection{ Normal density from phonon excitations and sound velocities}
To investigate the normal density and
sound velocities of the Hamiltonian~(\ref{eqn2}), it is necessary to  derive hydrodynamics
for low energy phonon excitation. Our starting point is the microscopic equation of the order
parameter, i.e, the time-dependent Gross-Pitaevskii (GP) equation. We assume
the order parameter
\begin{equation*}
|\psi \rangle =\left(
\begin{array}{c}
\sqrt{n_{1}}e^{i\theta _{1}} \\
\sqrt{n_{2}}e^{i\theta _{2}}%
\end{array}%
\right) ,
\end{equation*}%
which satisfies the time-dependent GP equation \cite{Zhengwei2012}.
Near the ground state, we expand the GP equations in terms of
small fluctuations $\delta n_s$ and $\delta \theta s$ and get four linear
equations
\begin{eqnarray*}
\partial _{t}\delta n_{1} &=&-\left[(p_{0}-k_{0})\partial _{x}\delta n_{1}+\bar{n}%
_{1}\nabla ^{2}\delta \theta _{1}\right] \\
&&+\Omega \sqrt{\bar{n}_{1}\bar{n}_{2}}(\delta \theta _{1}-\delta \theta
_{2}),\notag\\
\partial _{t}\delta n_{2} &=&-\left[(p_{0}+k_{0})\partial _{x}\delta n_{2}+\bar{n}%
_{2}\nabla ^{2}\delta \theta _{2}\right] \\
&&-\Omega \sqrt{\bar{n}_{1}\bar{n}_{2}}(\delta \theta _{1}-\delta \theta
_{2}),
\\
-\partial _{t}\delta \theta _{1} &=&-\frac{\nabla ^{2}\delta n_{1}}{4\bar{n}%
_{1}}+(p_{0}-k_{0})\partial _{x}\delta \theta _{1}+(g\delta n_{1}+g\delta
n_{2}) \\
&&-\frac{\Omega }{4}\left( \frac{\delta n_{2}}{\sqrt{\bar{n}_{1}\bar{n}_{2}}}%
-\sqrt{\frac{\bar{n}_{2}}{\bar{n}_{1}^{3}}}\delta n_{1}\right) ,
\\
-\partial _{t}\delta \theta _{2} &=&-\frac{\nabla ^{2}\delta n_{2}}{4n_{2}}%
+(p_{0}+k_{0})\partial _{x}\delta \theta _{2}+(g\delta n_{2}+g\delta n_{1})
\\
&&-\frac{\Omega }{4}\left( \frac{\delta n_{1}}{\sqrt{\bar{n}_{1}\bar{n}_{2}}}%
-\sqrt{\frac{\bar{n}_{1}}{\bar{n}_{2}^{3}}}\delta \bar{n}_{2}\right) ,
\end{eqnarray*}%
where $\bar{n}_{1(2)}$ denotes its average value in the ground state.

Next we introduce the total density fluctuation $\delta n=\delta
n_{1}+\delta n_{2}$, the spin polarization $\delta S_{z}=\delta n_{1}-\delta
n_{2}$, the common phase $\delta \theta =(\delta \theta _{1}+\delta \theta
_{2})/2$, and the relative phase $\delta \theta _{R}=\delta
\theta _{1}-\delta \theta _{2}$. For low energy ($\omega_\textbf{q} \rightarrow 0$) and long wave length ($q\rightarrow 0$) fluctuations, we adiabatically eliminate the spin parts, i.e., $\delta \theta _{R}$ and $\delta S_{z}$. Therefore we get effective hydrodynamic equation for the total
density $\delta n$ and common phase $\delta \theta $ \cite{Qu2017},
i.e.,
\begin{align}
& \partial _{t}\delta n=-n\left[\frac{\partial _{x}^{2}\delta \theta }{z_{1}}%
+(\partial _{y}^{2}+\partial _{z}^{2})\delta \theta \right],  \notag
\label{hydrodynamic1} \\
& -\partial _{t}\delta \theta =g\delta n,
\end{align}%
where $n=\bar{n}_{1}+\bar{n}_{2}$ is the average particle density in the
ground state. $z_{1}=1/\cos ^{2}(2\theta )=1/\left[ 1-\Omega ^{2}/\left(
4k_{0}^{4}\right) \right] $ describes  the enhancements of effective masses for the
plane-wave phase and $z_{1}=1/(1-2k_{0}^{2}/\Omega )$ for the
zero-momentum phase. Near the phase transition point ($\Omega \rightarrow
2k_{0}^{2}$), $z_{1}\rightarrow \infty $. From Eq.~(\ref{hydrodynamic1}), we
get the energies for phonon excitations as
\begin{equation*}
\omega _{\pm \textbf{q}}=c(\hat{q})q,
\end{equation*}%
where the sound velocity $c(\hat{q})\equiv \sqrt{\cos ^{2}(\alpha )/z_{1}+\sin ^{2}(\alpha )}%
c_{0}$, $c_{0}=\sqrt{gn}=\sqrt{\partial p/\partial n}=\sqrt{n\partial \mu
/\partial n}$ with $\mu =gn-\Omega ^{2}/\left( 8k_{0}^{2}\right) $ for the
plane-wave phase and $\mu =gn+\left( k_{0}^{2}-\Omega \right) /2$ for the
zero-momentum phase \cite{Martone2012}. $\hat{q}=\mathbf{q}/q=\{\cos (\alpha
),\sin (\alpha )\cos (\phi ),\sin (\alpha )\sin (\phi )\}$, and $\alpha $ is the
angle between $\hat{q}$ and $x$-axis.
Taking the spatial derivatives of the second equation and identifying $%
g\delta n\rightarrow \delta\mu $ (deviations relative to the ground state values), Eq.~(\ref{hydrodynamic1}) becomes the linear equation~(\ref%
{linearzero}) with $z_{\perp }=z_{2}=z_{3}=1$.

From Eqs.~(\ref{normaldensity}) and (\ref{secondsound}), we get the normal
density, the first and second sound
velocities in spin-orbit coupled BEC as
\begin{align}\label{normal}
& \rho _{n}(\hat{x})=z_{1}^{\frac{3}{2}}\frac{2\pi ^{2}T^{4}}{45\hbar
^{3}c_{0}^{5}},  \notag \\
& \rho _{n}(\hat{y})=\rho _{n}(\hat{z})=z_{1}^{\frac{1}{2}}\frac{2\pi
^{2}T^{4}}{45\hbar ^{3}c_{0}^{5}},  \notag \\
& c_{1}(\hat{q})=c_{0}\sqrt{\cos ^{2}(\alpha )/z_{1}+\sin ^{2}(\alpha )},
\notag \\
& c_{2}(\hat{q})=\frac{1}{(z_{1})^{1/4}}\frac{c_{1}(\hat{q})}{\sqrt{3}}.
\end{align}%
Along the $x$-direction, the corrections of the normal density and the second sound velocity are given by
\begin{align}\label{normal2}
&\beta_1=z_{1}^{\frac{3}{2}},\notag\\
&\gamma_1=1/z_{1}^{\frac{3}{4}}.
\end{align}%
From Eqs.~(\ref{normal}) and (\ref{normal2}), we see that with increasing the effective mass,
the normal density increases; while the second sound velocity decreases.
Especially, when $\Omega \rightarrow 2k_{0}^{2}$, i.e., near the phase transition point (%
$z_{1}=1/\sqrt{1-\Omega ^{2}/\left( 4k_{0}^{4}\right) }$ or $%
1/[1-2k_{0}^{2}/\Omega ]\rightarrow \infty $), the effective mass diverges along the $%
x$-axis direction. The normal density $\rho _{n}(\hat{x})$ from phonon
excitations would increase greatly (see Fig.~1),
while the second sound velocity along the $x$-axis direction $c_{2}(\hat{x}%
)\rightarrow 0$.
\begin{figure}[h]
\begin{center}
\includegraphics[width=\columnwidth]{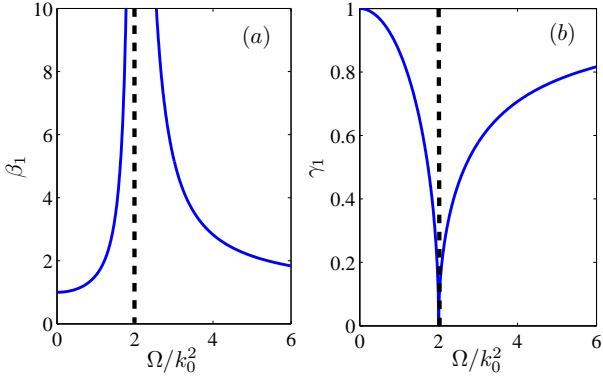}
\end{center}
\caption{ The corrections of the normal density [panel (a)] and
the second sound velocity [panel (b)] in spin-orbit coupled BEC (along the $x$-axis
direction). Note that near the phase transition ($\Omega
/k_{0}^{2}\rightarrow 2$), the effective mass would diverge, i.e., $%
z_{1}\rightarrow \infty $. }
\label{figure2}
\end{figure}

\subsection{ Superfluid density and Josephson relation}
With the above linearized hydrodynamic equations, in the following we give a unified derivation for superfluid density and Josephson relation in spin-orbit coupled BEC.
We should stress that one can take two different viewpoints on the effects
of enhancement of effective mass in Eq.~(\ref{hydrodynamic1}). The first one
is that the particle number density does not change, while the superfluid velocity
decreases due to a factor $1/z_{1}$, which is adopted by previous sections in
this paper. The another one is that the superfluid density decreases, while the superfluid velocity does not change, which would be adopted in following parts in this subsection.

We introduce superfluid density along the $\hat{q}$-direction $\rho _{s}(%
\hat{q})$ as
\begin{align}
& \rho _{s}(\hat{x})=n/z_{1},~\rho _{s}(\hat{y})=\rho _{s}(\hat{z})\equiv
\rho _{s\perp }=n,  \notag \\
& \rho _{s}(\hat{q})=\rho _{sx}\cos ^{2}(\alpha )+\rho _{s\perp }\sin
^{2}(\alpha ).
\end{align}%
In this case, Eq.~(\ref{hydrodynamic1}) becomes
\begin{align}
& \partial _{t}\delta n=-\rho _{s}(\hat{x})\partial _{x}^{2}\delta \theta
-\rho _{s\perp }(\partial _{y}^{2}+\partial _{z}^{2})\delta \theta ,  \notag
\label{hydrodynamic} \\
& -\partial _{t}\delta \theta =g\delta n.
\end{align}%
In the following, we will show that $\rho _{s}(\hat{q})$ is indeed the
superfluid density.

We can write down an effective Hamiltonian for the
hydrodynamic equation (\ref{hydrodynamic}) as
\begin{eqnarray}
H_{\text{eff}} &=&\frac{1}{2}\int d^{3}\mathbf{r}\{\rho _{s}(\hat{x}%
)(\partial _{x}\delta \theta )^{2}  \notag  \label{fluidhamiltonian} \\
&&+\rho _{s\perp }\left[(\partial _{y}\delta \theta )^{2}+(\partial _{z}\delta
\theta )^{2}\right]+g(\delta n)^{2}\}.
\end{eqnarray}%
Assuming the commutator relation $\left\{ \delta \theta (\mathbf{r}%
),\delta n(\mathbf{r^{\prime }})\right\} =-\delta ^{3}(\mathbf{r}-\mathbf{r}%
^{\prime })$ holds (Poisson brackets), we can easily get the above hydrodynamic
equation (\ref{hydrodynamic}) from the Hamilton's equations, i.e.,
\begin{equation*}
\partial _{t}\delta n(\mathbf{r})=\{\delta n(\mathbf{r}),H_{\text{eff}%
}\},\partial _{t}\delta \theta (\mathbf{r})=\{\delta \theta (\mathbf{r}),H_{%
\text{eff}}\}.
\end{equation*}%
Further assuming the quantized commutator relation $\left[\delta \theta (\mathbf{r}%
),\delta n(\mathbf{r}^{\prime })\right]=-i\delta ^{3}(\mathbf{r}-\mathbf{r}%
^{\prime })$ holds \cite{Lifshitz}, the phase $\delta \theta $ and density $\delta
n$ can be be expressed in terms of  the phonon's annihilation and creation operators
\begin{equation*}
\delta \theta (\mathbf{r},t)\!\!=\!\!\sum_{\mathbf{q}}\left[A_{\mathbf{q}}C_{%
\mathbf{q}}e^{i(\mathbf{q}\cdot \mathbf{r}-\omega _{\mathbf{q}}t)}+A_{%
\mathbf{q}}^{\ast }C_{\mathbf{q}}^{\dagger }e^{-i(\mathbf{q}\cdot \mathbf{r}%
-\omega _{\mathbf{q}}t)}\right],
\end{equation*}%
\begin{equation*}
\delta n(\mathbf{r},t)\!\!=\!\!\sum_{\mathbf{q}}\left[B_{\mathbf{q}}C_{\mathbf{q}%
}e^{i(\mathbf{q}\cdot \mathbf{r}-\omega _{\mathbf{q}}t)}+B_{\mathbf{q}%
}^{\ast }C_{\mathbf{q}}^{\dagger }e^{-i(\mathbf{q}\cdot \mathbf{r}-\omega _{%
\mathbf{q}}t)}\right],
\end{equation*}%
where $A_{\mathbf{q}}(B_{\mathbf{q}})$ is a coefficient to be determined and $C_{%
\mathbf{q}}$ is the annihilation operator for phonon. From the continuity
equation
\begin{equation*}
\partial _{t}\delta n=-\rho _{s}(\hat{x})\partial _{x}^{2}\delta \theta
-\rho _{s\perp }(\partial _{y}^{2}+\partial _{z}^{2})\delta \theta ,
\end{equation*}%
we get
\begin{equation*}
-ic(\hat{q})B_{q}=q\rho _{s}(\hat{q})A_{q}.
\end{equation*}%
From the commutator relation $\left[\delta \theta (\mathbf{r}),\delta n(\mathbf{r}%
^{\prime })\right]=-i\delta ^{3}(\mathbf{r}-\mathbf{r}^{\prime })$, we get $A_{%
\mathbf{q}}B_{\mathbf{q}}^{\ast }=-i/2$ and then $A_{\mathbf{q}}=\sqrt{c(%
\hat{q})/\left[ 2\rho _{s}(\hat{q})q\right] }$, $B_{\mathbf{q}}=i\sqrt{\rho
_{s}(\hat{q})q/\left[ 2c(\hat{q})\right] }$. Finally, we have
\begin{align}
&\delta \theta (\mathbf{r},t)\!\!=\!\!\sum_{\mathbf{q}}\sqrt{\frac{c(\hat{q})%
}{2\rho _{s}(\hat{q})q}}\left[C_{\mathbf{q}}e^{i(\mathbf{q}\cdot \mathbf{r}%
-\omega _{q}t)}+C_{\mathbf{q}}^{\dagger }e^{-i(\mathbf{q}\cdot
\mathbf{r}-\omega _{\mathbf{q}}t)}\right],\notag\\
&\delta n(\mathbf{r},t)\!\!=\!\!i\sum_{\mathbf{q}}\sqrt{\frac{\rho _{s}(\hat{q%
})q}{2c(\hat{q})}}\left[C_{\mathbf{q}}e^{i(\mathbf{q}\cdot \mathbf{r}-\omega
_{q}t)}-C_{\mathbf{q}}^{\dagger }e^{-i(\mathbf{q}\cdot \mathbf{r}%
-\omega _{\mathbf{q}}t)}\right].\label{zz}
\end{align}%
From Eq.~(\ref{zz}), we get density and phase fluctuations
in terms of phonon's operators as
\begin{align}
& n_{\mathbf{q}}=i\sqrt{\frac{\rho _{s}(\hat{q})q}{2c(\hat{q})}}\left( C_{%
\mathbf{q}}-C_{-\mathbf{q}}^{\dagger }\right) ,  \notag \\
& \theta _{\mathbf{q}}=\sqrt{\frac{c(\hat{q})}{2\rho _{s}(\hat{q})q}}\left(
C_{\mathbf{q}}+C_{-\mathbf{q}}^{\dagger }\right) .\label{nn}
\end{align}%
From Eq.~(\ref{nn}), we can verify that $\rho _{s}(\hat{q})$
is indeed the superfluid density. For example, the superfluid density can be
written as \cite{normaldensity}
\begin{eqnarray}\label{normalden}
\rho _{s}(\hat{q}) &=&c^{2}(\hat{q})\kappa (\hat{q})\notag \\
&=&\lim_{q\rightarrow 0}\frac{|\langle \mathbf{q}|n_{-\mathbf{q}%
}|0\rangle |^{2}\omega _{\mathbf{q}}+|\langle -\mathbf{q}|n_{\mathbf{q}%
}|0\rangle |^{2}\omega _{\mathbf{q}}}{q^{2}},
\end{eqnarray}%
where
\begin{equation*}
\kappa (\hat{q})=\lim_{q\rightarrow 0}\left[ \frac{|\langle \mathbf{q%
}|n_{\mathbf{q}}^{\dagger }|0\rangle |^{2}}{c(\hat{q})q}+\frac{|\langle -%
\mathbf{q}|n_{-\mathbf{q}}^{\dagger }|0\rangle |^{2}}{c(\hat{q})q}\right]
\end{equation*}%
is the compressibility, $|0\rangle $ is the ground state, and $|\mathbf{q}%
\rangle =C_{\mathbf{q}}^{\dagger }|0\rangle $ is the single-phonon state. In Eq.~(\ref{normalden}), we have used the fact that the single-phonon's
contribution is dominant in the compressibility \cite{normaldensity} and $%
\omega _{\pm\mathbf{q}}=c(\hat{q})q$. Due to influences of upper branch, in
spin-orbit coupled BEC, the superfluid density $\rho _{s}(\hat{q})$ would be
smaller than the total density, i.e., $\rho _{s}(\hat{q})<n$ \cite{normaldensity}%
. In this sense, we can interpret that the suppression of superfluid density in
spin-orbit coupled BEC is due to the enhancement of effective mass.


On the other hand, as $q\rightarrow 0$ and at low energy, the boson field
operator can be written as \cite{Lifshitz}
\begin{equation*}
\psi _{\sigma }(\mathbf{r})=\langle \psi _{\sigma }\rangle e^{i\delta \theta
(\mathbf{r})}\simeq \langle \psi _{\sigma }\rangle \left[ 1+i\delta \theta (%
\mathbf{r})\right].
\end{equation*}%
So we get
\begin{equation}
\psi _{\sigma ,\mathbf{q}}=i\langle \psi _{\sigma }\rangle \theta _{\mathbf{q%
}}=i\langle \psi _{\sigma }\rangle \sqrt{\frac{c(\hat{q})}{2\rho _{s}(\hat{q}%
)q}}\left[C_{\mathbf{q}}+C_{-\mathbf{q}}^{\dagger }\right].  \label{matrixelement}
\end{equation}%
With Eq.~(\ref{matrixelement}), the matrix element $\langle 0|\psi _{\sigma ,%
\mathbf{q}}|\mathbf{q}\rangle =i\langle \psi _{\sigma }\rangle \sqrt{\frac{c(%
\hat{q})}{2\rho _{s}(\hat{q})q}}$, $\langle -\mathbf{q}|\psi _{\sigma ,%
\mathbf{q}}|0\rangle =i\langle \psi _{\sigma }\rangle \sqrt{\frac{c(%
\hat{q})}{2\rho _{s}(\hat{q})q}}$, and the Green's function matrix $G_{\sigma,\sigma}(\mathbf{q},0)=-\sum_n\left[\frac{\langle0|\psi_{\sigma \textbf{q}}|n\rangle\langle n|\psi^{\dag}_{\sigma \mathbf{q}}|0\rangle}{\omega_{n0}}+\frac{\langle0|\psi^{\dag}_{\sigma \mathbf{q}}|n\rangle\langle n|\psi_{\sigma \mathbf{q}}|0\rangle}{\omega_{n0}}\right]\simeq-\frac{|\langle \psi _{\sigma }\rangle
|^{2}}{\rho _{s}(\hat{q})q^{2}}$  as $q\rightarrow 0$ \cite{zhangyicai2018}. In the above derivations, we have also used the fact that the single-phonon states have dominant contributions in the Green's function as $\mathbf{q}\rightarrow0$ and  excitation energies for single-phonon states
$\omega_{n0}=\omega _{\pm\mathbf{q}}=c(\hat{q})q$.
Using $n_{0}=\sum_{\sigma={1,2}} |\langle \psi
_{\sigma }\rangle |^{2}$, the Josephson relation is obtained \cite{zhangyicai2018}
\begin{equation}\label{Joseph}
\rho _{s}(\hat{q})=-\lim_{q\rightarrow 0}\frac{n_{0}}{q^{2}tr G(%
\mathbf{q},0)}.
\end{equation}%
The superfluid density from the Josephson relation [Eq.~(\ref{Joseph})] is also consistent with the current-current
correlation calculations \cite{normaldensity}.

From Eq. (\ref{matrixelement}) of $\psi _{\sigma ,\textbf{q}}$, we get the momentum
distribution function as $q\rightarrow 0$,
\begin{equation*}
N_{\mathbf{q}}=\sum_{\sigma=1,2 }\langle \psi _{\sigma ,\mathbf{q}}^{\dagger
}\psi _{\sigma ,\mathbf{q}}\rangle =\frac{n_{0}c(\hat{q})}{2\rho _{s}(\hat{q}%
)q}\left( 2n_{\mathbf{q}}+1\right) ,
\end{equation*}%
where $n_{\mathbf{q}}=1/\left( e^{\omega _{\mathbf{q}}/T}-1\right)$ %
 is the phonon Bose distribution function for the rest frame. Specially, at $T=0$ and as $%
q\rightarrow 0$, $N_{\mathbf{q}}=n_{0}c(\hat{q})/\left[ 2\rho _{s}(\hat{q})q%
\right] \propto 1/q$; when $\omega _{\mathbf{q}}\ll T$, $N_{\mathbf{q}%
}=n_{0}T/\left[ \rho _{s}(\hat{q})q^{2}\right] \propto 1/q^{2}$, which are generalizations of the isotropic results \cite%
{Bogoliubov,Lifshitz}.

Using the effective Hamiltonian~(\ref{fluidhamiltonian}), we can
calculate the phase or density fluctuations within the hydrodynamic
formalism \cite{Pitaevskii}. The energy in the momentum space is given by
\begin{align}
& \delta E=H_{\text{eff}}=\frac{1}{2}\int d^{3}\mathbf{r}\left\{ \rho _{s}(%
\hat{x})(\partial _{x}\delta \theta )^{2}\right.   \notag \\
& +\left. \rho _{s\perp }\left[(\partial _{y}\delta \theta )^{2}+(\partial
_{z}\delta \theta )^{2}\right]+g(\delta n )^{2}\right\}   \notag \\
& =\frac{1}{2}\sum_{\mathbf{q}}\left[\rho _{s}(\hat{q})q^{2}|\theta _{\mathbf{q}%
}|^{2}+g|n_{\mathbf{q}}|^{2}\right].
\end{align}%
Because the thermal probability distribution
\begin{equation*}
P\propto e^{-\delta E/T}=e^{-\frac{1}{2}\sum_{\mathbf{q}}\left[\rho _{s}(\hat{q}%
)q^{2}|\theta _{\mathbf{q}}|^{2}+g|n_{\mathbf{q}}|^{2}\right]/T},
\end{equation*}%
for the long wave lengths ($\omega _{\mathbf{q}}\ll T$), the
thermal fluctuations of the phase and density are given by
\begin{equation*}
\langle |\theta _{\mathbf{q}}|^{2}\rangle =\frac{T}{\rho _{s}(\hat{q})q^{2}}%
\text{, \ }\langle |n_{\mathbf{q}}|^{2}\rangle =\frac{T}{g}.
\end{equation*}%
Along the $x$-axis direction ($\hat{q}=\hat{x}$), we see the phase fluctuation near phase transition point [$\rho _{s}(\hat{x})\rightarrow 0$] is
very dramatic and diverges, while the density fluctuation is always finite.
\newline

\section{Conclusion}

In summary, we have generalized the two-fluid theory to a superfluid system with anisotropic effective masses. As a specific example, this theory is used to investigate spin-orbit coupled BEC realized in recent experiments. At low temperature, the normal density from phonon excitations and the second sound velocity have been obtained analytically. Near the phase transition from the plane wave to zero-momentum phases, due to the effective mass divergence, the normal density from phonon excitation increases greatly, while the second sound velocity is suppressed significantly. With quantum hydrodynamic formalism, we have given a unified derivation for the suppressed superfluid density and Josephson relation.

Before ending up this paper, we make three remarks. The first is that our previous calculations are restricted to the case of $z>1$. However, our theory can be extended straightforwardly to the other case of $0<z<1$. The main results are similar and thus are not discussed here. The second is that for the spin-coupled BEC at higher temperature, the up gapped excitation would play an important role in hydrodynamics. Thus, how to take account of the up branch excitations properly and construct corresponding hydrodynamic theory still needs further investigations. The last is that when the system is exactly at the
phase-transition point from the plane-wave and zero momentum phases, the quadratic effective mass terms ($\propto p^2$) in the Hamiltonian (\ref{TH}) would vanish, while quartic terms ($\propto p^4$) may play an important role. In such case, the corresponding hydrodynamics also needs further investigations.

\section{Acknowledgements}

Yi-Cai Zhang thank Shizhong Zhang for useful discussions. This work was
supported by the NSFC under Grants No.~11874127, No.~11674200, No.~11747079,
No.~61565007, No.~11875149, No.~61565013, No.~11434015 and No.~61835013, the
National Key R\&D Program of China under grants No.~2016YFA0301500, SPRPCAS
under grants No.~XDB01020300 and No.XDB21030300, and Hong Kong Research
Grants Council (General Research Fund, Grant No.~HKU 17318316 and
Collaborative Research Fund, Grant No.~C6026-16W).
 Yi-Cai Zhang also
acknowledge the support of a startup grant from Guangzhou University.

\end{document}